\documentclass[a4paper]{jpconf}
\usepackage{graphicx}
\usepackage{cite}
\usepackage{amsmath}
\usepackage{amssymb}

\def\ep{\varepsilon}
\def\cM{\cal M}
\def\cI{\cal I}

\begin{document}
\title{{\footnotesize  \sf  DESY 19--078, DO-TH 19/06 \hfill}\\
Three loop QCD corrections to heavy quark form factors}

\author{J.~Ablinger$^1$, J.~Bl\"umlein$^2$, P.~Marquard$^2$, N.~Rana$^{2,3}$, C.~Schneider$^1$}

\address{$^1$RISC, Johannes Kepler University, Altenbergerstra{\ss}e 69, A-4040, Linz, Austria}
\address{$^2$DESY, Platanenallee 6, D-15738 Zeuthen, Germany}
\address{$^3$INFN, Sezione di Milano, Via Celoria 16, I-20133 Milano, Italy}

\ead{Narayan.Rana@mi.infn.it}

\begin{abstract}
Higher order calculations in perturbative Quantum Field Theories often produce coupled linear systems 
of differential equations which factorize to first order. Here we present an algorithm to solve
such systems in terms of iterated integrals over an alphabet the structure of which is implied by 
the coefficient matrix of the given system. We apply this method to calculate the master integrals in the 
color--planar and complete light quark contributions to the three-loop massive form factors.
\end{abstract}

\section{Introduction}
Higher order calculations in perturbative Quantum Field Theories often produce coupled linear systems 
of differential equations which factorize to first order. 
Through the integration-by-parts identities (IBP) \cite{Chetyrkin:1981qh,Laporta:2001dd}, the Feynman integrals 
appearing in a scattering amplitude can be mapped to a smaller set of integrals, the master integrals (MIs).
One method to solve these integrals is the method of differential equations 
\cite{Kotikov:1990kg,Remiddi:1997ny,Henn:2013pwa,Ablinger:2015tua}, where one differentiates the MIs with respect to a 
kinematic invariant and obtains such a coupled linear system of differential equations in which we are interested in.
Such an algorithm was constructed earlier in \cite{Ablinger:2015tua, Blumlein:2017dxp} in the case of a system of 
difference equations. Here we summarize the algorithm, presented in \cite{Ablinger:2018zwz}, operating on uni-variate 
systems of differential equations, 
which factorize to first order.  
In the cases where the systems do not factorize to first order, elliptic and even more involved structures appear,
 cf. e.g. \cite{Laporta:2004rb,Bloch:2013tra,Adams:2015gva,Adams:2014vja,Adams:2016xah,Ablinger:2017bjx,Broedel:2017kkb,
Broedel:2018rwm,BOOK}. Our algorithm works for any basis of master integrals, while in the one in \cite{Henn:2013pwa} 
a special basis is required.

As an example, we employ this method to compute the set of MIs which appear in the color--planar and complete light quark 
non–singlet three-loop contributions to the heavy-quark form factors for different currents, namely the vector, 
axial-vector, scalar and pseudo--scalar currents.
These form factors are of phenomenological importance. Being the heaviest particle of the Standard Model (SM), the top 
quark plays an important role in comprehending the electro-weak symmetry breaking (EWSB). 
Consequently, many observables like the inclusive production cross section of a pair of top quarks, the
forward--backward asymmetry, etc., draw attention in the post era of the Higgs boson discovery.  
These form factors are basic building blocks of many such observables.
The two-loop QCD contributions to these massive form factors were first computed in 
\cite{Bernreuther:2004ih,Bernreuther:2004th,Bernreuther:2005rw,Bernreuther:2005gw}.  
Later in \cite{Gluza:2009yy}, an independent computation was performed for the vector form factors, including 
the ${\cal O}(\ep)$ terms, where $\ep$ is the dimensional regularization parameter in $D=4-2\ep$ space-time dimensions.

In \cite{Ablinger:2017hst}, the two-loop QCD contributions up to ${\cal O}(\ep^2)$ for all massive form 
factors were calculated.  At the three-loop level, the color-–planar contributions to the vector form factors have been 
computed in \cite{Henn:2016tyf,Henn:2016kjz} and the complete light quark contributions in \cite{Lee:2018nxa}.
Using the method described here, we have calculated both the color–-planar and complete light quark contributions to 
the three--loop form factors for the axial--vector, scalar and pseudo-scalar currents in \cite{Ablinger:2018yae}
and for the vector current in \cite{Ablinger:2018zwz}, where a detailed description of the present method along with 
an example is presented. In a parallel calculation, the same results have been obtained in \cite{Lee:2018rgs}. 
The asymptotic behavior of these three--loop form factors has been studied in \cite{Blumlein:2018tmz,Ahmed:2017gyt}.

\section{Description of the method}

We consider a set of $n$ MIs ${\cal I} = (I_1,\ldots,I_n)$ within the same topology. The MIs are 
functions of the space--time dimension $D$ and the Landau variable $x$ defined by
\begin{equation}
\label{EQ:1}
s = \frac{q^2}{m^2} = -\frac{(1-x)^2}{x}.
\end{equation}
Here $q^2$ is the virtuality of the boson and $m$ denotes the heavy quark mass.
One obtains an $n \times n$ system of coupled linear differential equations by performing the derivative with respect to 
$x$ of each of the MIs followed by the IBP reduction,
\begin{equation}\label{Equ:InputSystem}
 \frac{d}{dx} {\cal I} = {\cal M ~ I + R}.
\end{equation}
Here the $n\times n$ matrix ${\cal M}$ consists out of entries from the rational function field ${\mathbb K}(D,x)$ 
(or equivalently from ${\mathbb K}(\ep,x)$) where ${\mathbb K}$ is a field of characteristic $0$. 
The inhomogeneous part ${\cal R}=({\cal R}_1,\dots,{\cal R}_n)$ contains MIs which are already known. In simple 
cases, ${\cal R}$ turns out to be just the null vector. 
For more involved cases, we assume that each entry ${\cal R}_i$
is expanded\footnote{In the following $f^{(k)}$ does not denote the $k$th derivative of $f$.}
into a Laurent series in $\ep$ in terms of iterated integrals
\begin{equation}
{\cal R}_i = \sum_{j=-k}^{\infty} \ep^j {\cal R}_i^{(j)} \,.
\end{equation}
To proceed, we assume that 
the unknown integrals ${\cal I}_i$ can be expanded into a Laurent series in $\ep$
\begin{equation}\label{Equ:ExpansionIR}
{\cal I}_i = \sum_{j=-k}^{\infty} \ep^j {\cal I}_i^{(j)}.
\end{equation}
We first apply Z\"urcher's algorithm 
\cite{Zuercher:94,BCP13,NewUncouplingMethod,CoupledSys:15} (or a variant of it), implemented in the package {\tt OreSys} 
\cite{ORESYS}, leading to a single differential equation which is then analyzed with {\tt HarmonicSums} 
\cite{Vermaseren:1998uu,Blumlein:1998if,
Ablinger:2014rba, Ablinger:2010kw, Ablinger:2013hcp, Ablinger:2011te, Ablinger:2013cf,
Ablinger:2014bra,Ablinger:2017Mellin} 
for first order factorization.

The MIs can be distinguished sector--wise, where 
the maximal set of non--vanishing propagators in a single Feynman graph defines a sector
and corresponding subsets define the sub-sectors.
On the other hand, a derivative can only introduce the inverse of a new propagator.
Hence, the differential equation of a MI can contain integrals from the same sector or its sub--sectors.
Thus, organizing the MIs in such a way that MIs with less number of 
propagators are kept at the bottom of the list, provides an upper--block--triangular form of ${\cal M}$,
i.e. the diagonal elements of ${\cal M}$ are square matrices of rank one or more.
Each such square matrix represents a completely coupled set of MIs and we call them sub--systems of 
${\cal M}$. 
Due to such an arrangement of the system, we can now solve it in a bottom--up approach,
\textit{i.e.} we first solve the last coupled sub--system, having no dependence to the other MIs and work upwards in the 
next step.
Below we present how we solve the sub--systems.

\vspace*{1mm} \noindent
{\bf I.}~Let us consider $m$ integrals ${\cal \tilde I}=(\tilde I_1,\ldots, \tilde I_m)$ 
which constitute a coupled sub--system, 
\begin{equation} \label{eq:detilde}
 \frac{d}{dx} {\cal \tilde I} = {\cal \tilde M ~ \tilde I + \tilde R}, 
\end{equation}
where ${\cal \tilde M}$ and ${\cal \tilde R}$ are respective parts of ${\cal M}$ and ${\cal R}$
and hence have similar expansions in $\ep$.
Now we exploit the fact that for a certain topology and kinematics
an integral has a definitive pole structure with a fixed order ($k$) of the highest pole,
as \textit{e.g.} the integrals arising in three--loop heavy quark form factors can have at most a $1/\ep^3$ pole. 
Keeping that in mind, we plug in Eq.~\eqref{Equ:ExpansionIR} into Eq.~\eqref{eq:detilde}, perform the series expansion in $\ep$ and obtain the 
coefficient of $\ep^{k}$ as follows
\begin{equation} \label{eq:detildek}
 \frac{d}{dx} {\cal \tilde I}^{(k)} = {\tilde \cM^{(0)} ~ \tilde \cI^{(k)} 
 + \Big( \tilde \cM^{(1)} ~ \tilde \cI^{(k-1)} + \tilde \cM^{(2)} ~ \tilde \cI^{(k-2)} + \cdots 
+ \tilde \cM^{(k+l)} ~ \tilde \cI^{(-l)} \Big) + \tilde {\cal R}^{(k)}} \,.
\end{equation}

\vspace*{2mm}
\noindent
{\bf II.}~Now we try to solve the sub--system order by order in $\ep$.
To accomplish that we start with the coefficient of the leading pole 
$\ep^{-l}$. The corresponding sub--system is
\begin{equation} \label{eq:detildeml}
 \frac{d}{dx} {\cal \tilde I}^{(-l)} = {\tilde \cM^{(0)} ~ \tilde \cI^{(-l)} + \tilde {\cal R}^{(-l)}} \,. 
\end{equation}
To solve Eq.~(\ref{eq:detildeml}), a natural first step is to reduce 
this $m \times m$ system to a higher order linear differential equation. We refer 
to this procedure as `uncoupling'. By using {\tt OreSys}, we obtain 
\begin{equation} 
\label{eq:MAST}
\sum_{k=0}^{m} p_k(x)\frac{d^k}{dx^k} {\cal \tilde I}^{(-l)}_1(x) = r(x).
\end{equation}
Here $p_l(x)$ are rational functions in ${\mathbb K}(x)$ and $r(x)$ consists of contributions from ${\cal R}^{(-l)}_j(x)$
and its derivatives.
{\tt OreSys} also provides the additional $m-1$ relations for the remaining integrals ${\cal \tilde I}^{(-l)}_k(x)$, 
$k=2,\ldots,m$
in terms of linear combinations of ${\cal \tilde I}^{(-l)}_1(x)$ and its derivatives:
\begin{equation}\label{Equ:LinComb}
{\cal \tilde I}^{(-l)}_k(x)=\sum_{i=0}^{m-1}a_{k,i}(x)\,\frac{d^i}{dx^i}{\cal \tilde I}^{(-l)}_1(x)+\rho_k(x),
\end{equation}
with $a_{k,i} (x) \in {\mathbb K}(x)$ and $\rho_k (x) \in {\mathbb K}(x)$. 
We remark that the uncoupling may find several linear differential equations for several unknowns.
However, the sum of the order of all the linear differential equations is $m$.
For simplicity we assume in the following that only one linear differential equation for ${\cal \tilde I}^{(-l)}_1(x)$ is produced.

\vspace*{1mm}
\noindent
{\bf III.} Now we consider the homogeneous 
solutions of Eq.~(\ref{eq:MAST}). The differential equation can be factorized at first order as
\begin{equation}
\label{eq:FACT}
\left(\frac{d}{dx} - \hat{p}_1(x)\right)
\left(\frac{d}{dx} - \hat{p}_2(x)\right) 
{\dots} 
\left(\frac{d}{dx} - \hat{p}_m(x)\right) y_1(x) = 0,
\end{equation}
with $\hat{p}_k$ being rational functions in ${\mathbb K}(x)$
by using algorithms from~\cite{Singer:91,Bronstein:92,Hoeij:97}.
We find the d'Alembertian solutions $y_i(x)$, $i=1,\ldots,m$ (see \cite{Ablinger:2018zwz} for details) for the homogeneous part of Eq.~(\ref{eq:MAST})
in terms of iterative integrals. In our application to the three--loop massive form factors, the alphabet is 
\begin{equation}\label{Equ:Alphabet}
\mathfrak{A} = \left\{\tfrac{1}{x},~\tfrac{1}{1-x},
~\tfrac{1}{1+x}, 
~\tfrac{1}{1+x^2}, 
~\tfrac{x}{1+x^2}, 
~\tfrac{1}{1+x+x^2},~\tfrac{x}{1+x+x^2},
~\tfrac{1}{1-x+x^2},~\tfrac{x}{1-x+x^2} \right\} \,.
\end{equation}
The arising iterative integrals in our application can be simplified to the
harmonic polylogarithms (HPLs) \cite{Remiddi:1999ew} and cyclotomic HPLs \cite{Ablinger:2011te}.
We also note here that to fix the boundary condition of the differential equations, one needs
to evaluate these iterative integrals at a particular value of $x$, say $x=1$ which resulted in the
example on hand into multiple zeta values (MZVs) \cite{Blumlein:2009cf}
and cyclotomic constants \cite{Broadhurst:1998rz,Kalmykov:2010xv,Ablinger:2011te,Ablinger:2017tqs}.
In \cite{Henn:2015sem}, {\tt PSLQ} \cite{PSLQ} was used to conjecture 
relations beyond those known from \cite{Broadhurst:1998rz,Kalmykov:2010xv,Ablinger:2011te,Ablinger:2017tqs}.

\vspace*{1mm}
\noindent
{\bf IV.} The general solution ($g(x)$) including the inhomogeneous part can now be easily given by the method of variation of constants
in terms of iterative integrals as
\begin{equation}\label{Equ:SecondParticularRep}
 g(x) = \sum_{i=1}^m y_i (x)\int_l^{x} d\tilde{x} \frac{r(\tilde{x}) W_i(\tilde{x})}{W(\tilde{x})}
\end{equation}
where $W(\tilde{x})$ is the Wronskian of the linear differential equation Eq.~(\ref{eq:MAST}) 
and $W_i(\tilde{x})$ is given by
\begin{equation}
W_i(x) = (-1)^{i+m} \left|
\begin{array}{cccccc}    
        y_1         & \hdots & y_{i-1}         & y_{i+1}         & \hdots & y_m\\
        \frac{d}{dx}y_1   & \hdots & \frac{d}{dx}y_{i-1}   & \frac{d}{dx}y_{i+1}   & \hdots & \frac{d}{dx}y_m\\
        \vdots      &        & \vdots          & \vdots          &        & \vdots    \\
        \frac{d^{m-2}}{dx^{m-2}}y_1 & \hdots & \frac{d^{m-2}}{dx^{m-2}}y_{i-1} & \frac{d^{m-2}}{dx^{m-2}}y_{i+1} & \hdots & \frac{d^{m-2}}{dx^{m-2}}y_m\\
\end{array} \right| \,. 
\end{equation} 

\vspace*{1mm}
\noindent
{\bf V.} 
Next we use the general solution $(g(x))$ for ${\cal \tilde I}^{(-l)}_1(x)$, plug it into Eq.~(\ref{Equ:LinComb}) for $k=2,\dots,m$
and obtain solutions for the rest of the integrals and obtain ${\cal \tilde I}^{(-l)} (x)$ at ${\cal O}(\ep^{-l})$.

\vspace*{1mm}
\noindent
{\bf VI.} Now we consider the next order ($-l+1$) in the $\ep$-expansion. We use the solution of ${\cal \tilde 
I}^{(-l)}(x)$ in terms of iterative integrals and plug it
into Eq.~(\ref{eq:detildek}) for $k=-l+1$. Thus we obtain a new system of the form Eq.~(\ref{eq:detildeml}) for the 
$\ep^{-l+1}$-coefficient ${\cal \tilde I}^{(-l+1)}=({\cal \tilde I}_1^{(-l+1)}(x),\dots,{\cal \tilde I}_m^{(-l+1)}(x))$. 
Then the above procedure is repeated.
From Eq.~(\ref{eq:detildek}) it is evident that the homogeneous solution at each order remains same. The only changes 
happen to 
the inhomogeneous functions $r(x)$. As a consequence, we can reuse the already available homogeneous solutions 
$y_1(x),\dots,y_m(x)$ and 
just need to compute $g(x)$ in Eq.~\eqref{Equ:SecondParticularRep} with the updated function $r(x)$.
In Ref.~\cite{Ablinger:2018zwz} the method has been illustrated by a detailed example.

\section{Application to the massive form factors} 

\noindent
We consider the decay of a virtual massive boson,
which can be a vector ($V$), an axial-vector ($A$), a scalar ($S$) or a pseudo-scalar ($P$)
of momentum $q$, into a pair of heavy quarks of mass $m$, momenta $q_1$ and $q_2$ and color
$c$ and $d$, through a vertex $X_{cd}$, where
$X_{cd} = \Gamma^{\mu}_{V,cd}, \Gamma^{\mu}_{A,cd}, \Gamma_{S,cd}$ and
$\Gamma_{P,cd}$.
The general form of the amplitudes can be written as
\begin{align}
\bar{u}_c (q_1) \Gamma_{V,cd}^{\mu} v_d (q_2) &\equiv
  -i \bar{u}_c (q_1) \Big[ \delta_{cd}
 v_Q \Big( \gamma^{\mu} ~F_{V,1} 
         + \frac{i}{2 m} \sigma^{\mu \nu} q_{\nu}  ~ F_{V,2}  \Big) \Big] v_d (q_2) , 
\nonumber\\
\bar{u}_c (q_1) \Gamma_{A,cd}^{\mu} v_d (q_2) &\equiv
  -i \bar{u}_c (q_1) \Big[ \delta_{cd}
 a_Q \Big( \gamma^{\mu} \gamma_5 ~F_{A,1} 
         + \frac{1}{2 m} q_{\mu} \gamma_5  ~ F_{A,2}  \Big) \Big] v_d (q_2) , 
\nonumber\\
\bar{u}_c (q_1) \Gamma_{S,cd} v_d (q_2) &\equiv
  -i \bar{u}_c (q_1) \Big[ \delta_{cd}
 s_Q \Big( \frac{m}{v} (-i) ~F_{S}  \Big) \Big] v_d (q_2) , 
\nonumber\\
\bar{u}_c (q_1) \Gamma_{P,cd} v_d (q_2) &\equiv
  -i \bar{u}_c (q_1) \Big[ \delta_{cd}
 p_Q \Big( \frac{m}{v} (\gamma_5) ~F_{P}  \Big) \Big] v_d (q_2) \,.
\end{align}
Here $\bar{u}_c (q_1)$ and $v_d (q_2)$ are the bi--spinors of the quark and the anti--quark, 
respectively, with $\sigma^{\mu\nu} = \frac{i}{2} [\gamma^{\mu},\gamma^{\nu}]$.
$v_Q, a_Q, s_Q$ and $p_Q$ are the 
Standard Model (SM) coupling constants for the vector, axial-vector, scalar and pseudo-scalar, respectively;
$v = (\sqrt{2} G_F)^{-1/2}$ denotes the SM vacuum expectation value of the Higgs field, with the Fermi
constant $G_F$. More details can be found in \cite{Ablinger:2017hst}.
The ultraviolet (UV) renormalized form factors ($F_I$) are expanded in the strong coupling constant $\alpha_s = g_s^2/(4\pi)$ as follows
\begin{equation}
 F_{I} = \sum_{n=0}^{\infty} \left( \frac{\alpha_s}{4 \pi} \right)^n F_{I}^{(n)} \,.
\end{equation}
The form factors can be obtained from the amplitudes by multiplying 
appropriate projectors as given in \cite{Ablinger:2017hst} and performing the trace over the color and spinor indices.
$n_l$ and $n_h$ are the numbers of light and heavy quarks, respectively.

The computational procedure of the three--loop massive form factors are outlined 
in Refs.~\cite{Ablinger:2017hst,Ablinger:2018yae}. As usual the packages {\tt QGRAF} \cite{Nogueira:1991ex}, {\tt 
Color} \cite{vanRitbergen:1998pn},
{\tt Q2e/Exp} \cite{Harlander:1997zb,Seidensticker:1999bb} and {\tt FORM} \cite{Vermaseren:2000nd, Tentyukov:2007mu}
have been used to generate the Feynman diagrams, calculate their color structure and
to perform traces over the Dirac matrices. 
Since we use dimensional regularization,
an appropriate description for the treatment of $\gamma_5$ is needed in the case of axial-vector and pseudo-scalar form 
factors. However, both the color--planar and complete $n_l$ contributions belong to the so-called non-singlet case, 
where the vertex is connected to open heavy quark lines.
Hence, both $\gamma_5$-matrices appear in the same chain of Dirac matrices, which allows us
to use an anti-commuting $\gamma_5$ in $D$ space-time dimensions, with $\gamma_5^2 = 1$. This is implied 
by the well-known Ward identity, 
\begin{equation} \label{eq:cwi}
 q^{\mu} \Gamma_{A,cd}^{\mu, \sf ns} = 2 m \Gamma_{P,cd}^{\sf ns} \,,
\end{equation}
The IBP reduction to MIs has been performed using {\tt Crusher} \cite{CRUSHER}.
Finally, we have obtained 109 MIs, out of which 96 appear in the color--planar case.
To obtain the MIs, we have implemented the method described in the previous section.
Apart from {\tt OreSys}, we have intensively used {\tt HarmonicSums}
\cite{Vermaseren:1998uu,Blumlein:1998if,
Ablinger:2014rba, Ablinger:2010kw, Ablinger:2013hcp, Ablinger:2011te, Ablinger:2013cf,
Ablinger:2014bra,Ablinger:2017Mellin} and {\tt Sigma} 
\cite{Schneider:2007a,Schneider:2013a}. Finally, we have performed numerical checks of all the MIs 
using {\tt FIESTA} \cite{Smirnov:2008py, Smirnov:2009pb, Smirnov:2015mct}.

Once, we compute the MIs including the required orders in $\ep$, we finally obtain the color--planar
and complete light quark contributions to the unrenormalized three--loop massive form factors.
In these cases, along with the strong coupling constant, the heavy quark mass and the heavy quark 
wave function need UV renormalization. We consider the $\overline{\rm MS}$ scheme to renormalize
the strong coupling constant, where we set the universal factor $S_\varepsilon = 
\exp(-\varepsilon (\gamma_E - \ln(4\pi))$ for each loop order to one at the end of the calculation.
However, we renormalize the heavy quark mass and wave function in the on--shell (OS) scheme. The 
required renormalization constants are denoted by 
$Z_{m, {\rm OS}}$ \cite{Broadhurst:1991fy, Melnikov:2000zc,Marquard:2007uj,
Marquard:2015qpa,Marquard:2016dcn}, 
$Z_{2,{\rm OS}}$ \cite{Broadhurst:1991fy, Melnikov:2000zc,Marquard:2007uj,Marquard:2018rwx} and 
$Z_{a_s}$ \cite{Tarasov:1980au,Larin:1993tp,vanRitbergen:1997va,Czakon:2004bu,Baikov:2016tgj,Herzog:2017ohr,
Luthe:2017ttg} for the heavy quark mass, wave function and strong coupling constant, respectively. 

The QCD amplitudes contain infrared (IR) divergences arising from soft gluons and collinear partons.
However, such behavior is universal and acts as a check of exact computations.
In the case of massive form factors, the structure of IR singularities are given by \cite{Mitov:2006xs,Becher:2009kw}
\begin{equation}
 F_{I} = Z (\mu) F_{I}^{\mathrm{fin}} (\mu)\, ,
\end{equation}
where $F_{I}^{\mathrm{fin}}$ is finite as $\ep \rightarrow 0$. $Z(\mu)$ is related 
to the massive cusp anomalous dimension \cite{Grozin:2014hna,Grozin:2015kna} through the renormalization group equation.

\section{Results}
The system of differential equations of the MIs appearing in the
color--planar and complete light quark contributions to the massive form factors 
is a single scale and first order factorizable system. We apply our proposed method 
to this system to obtain the solutions of the MIs up to the required order in $\ep$.
Using these solutions for the MIs, we then obtain the form factors for vector, axial--vector, 
scalar and pseudo--scalar currents in the corresponding scenario.
The results are lengthy and hence are provided as supplementary material along with 
Ref.~\cite{Ablinger:2018yae, Ablinger:2018zwz}.

\begin{figure}[ht]
\centerline{%
\includegraphics[width=0.49\textwidth]{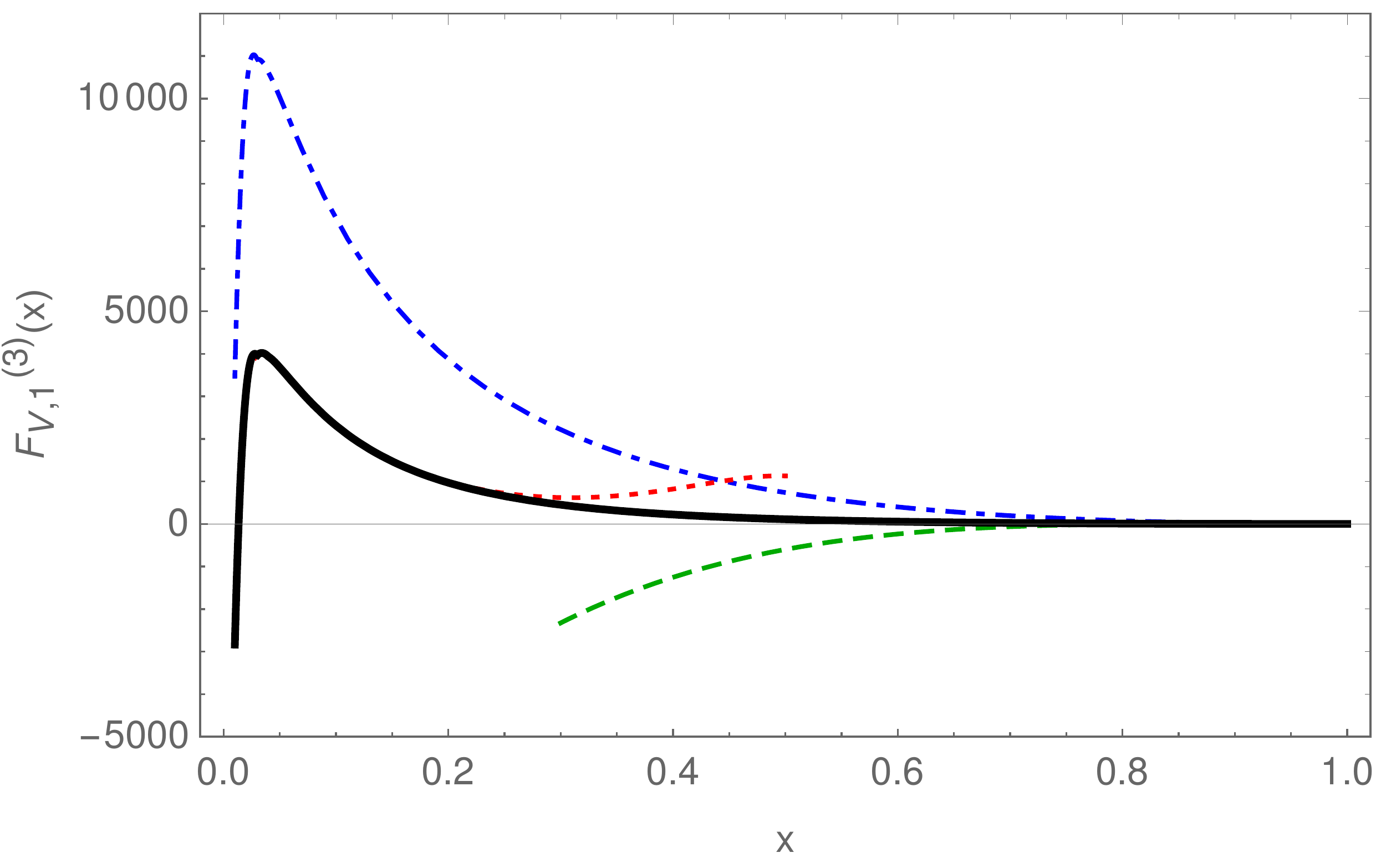}
\includegraphics[width=0.49\textwidth]{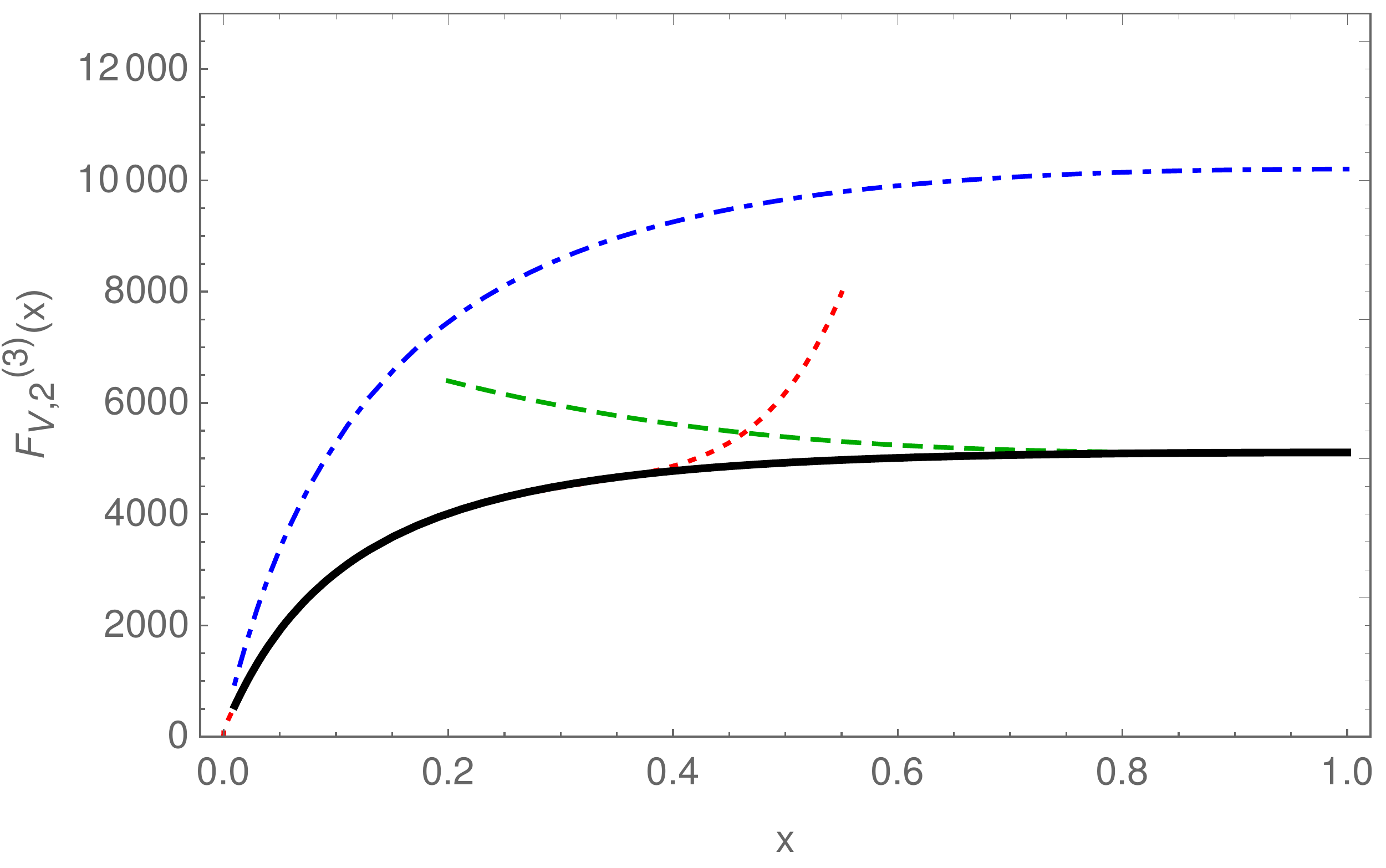}}
{
\includegraphics[width=0.49\textwidth]{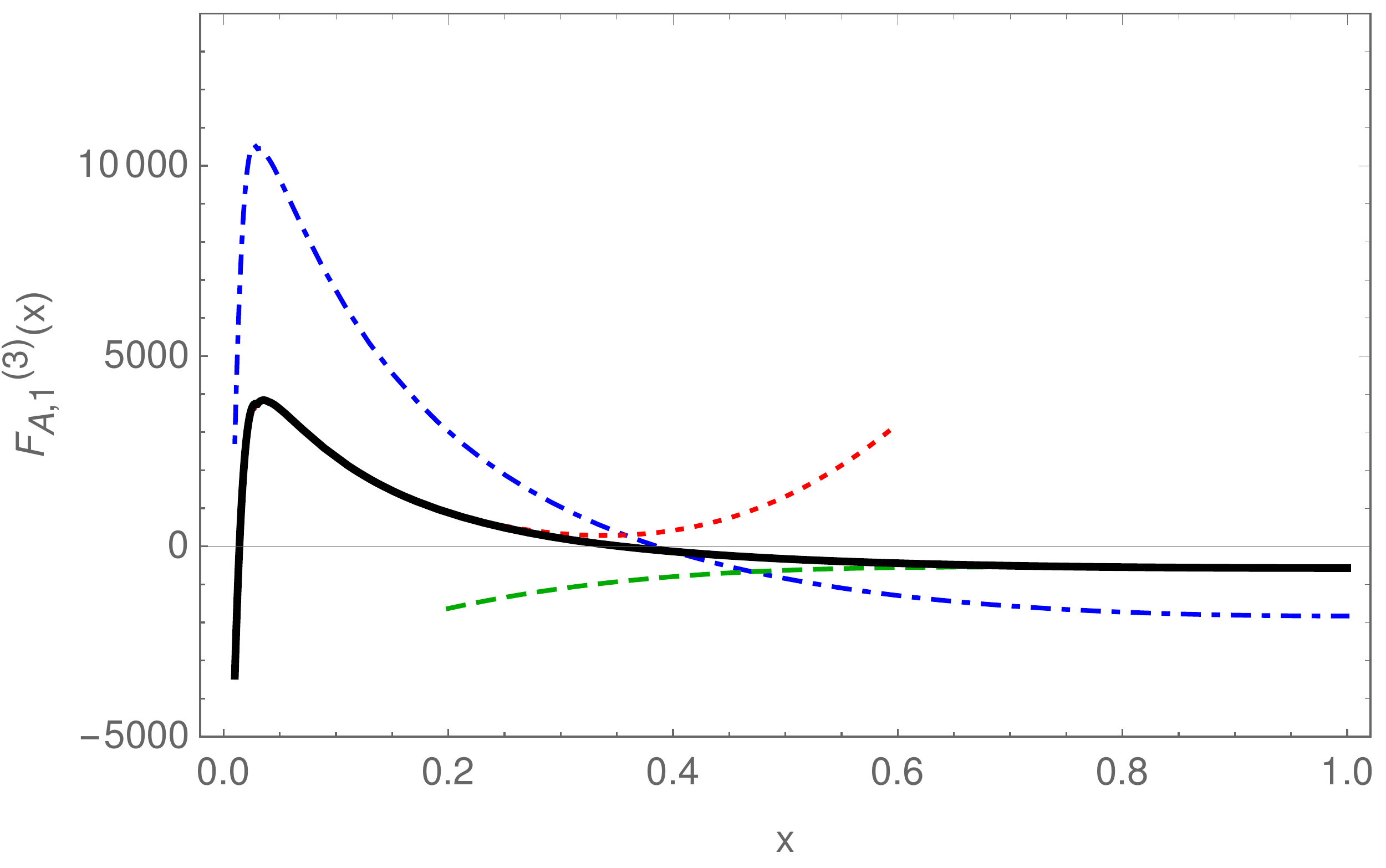}
\includegraphics[width=0.49\textwidth]{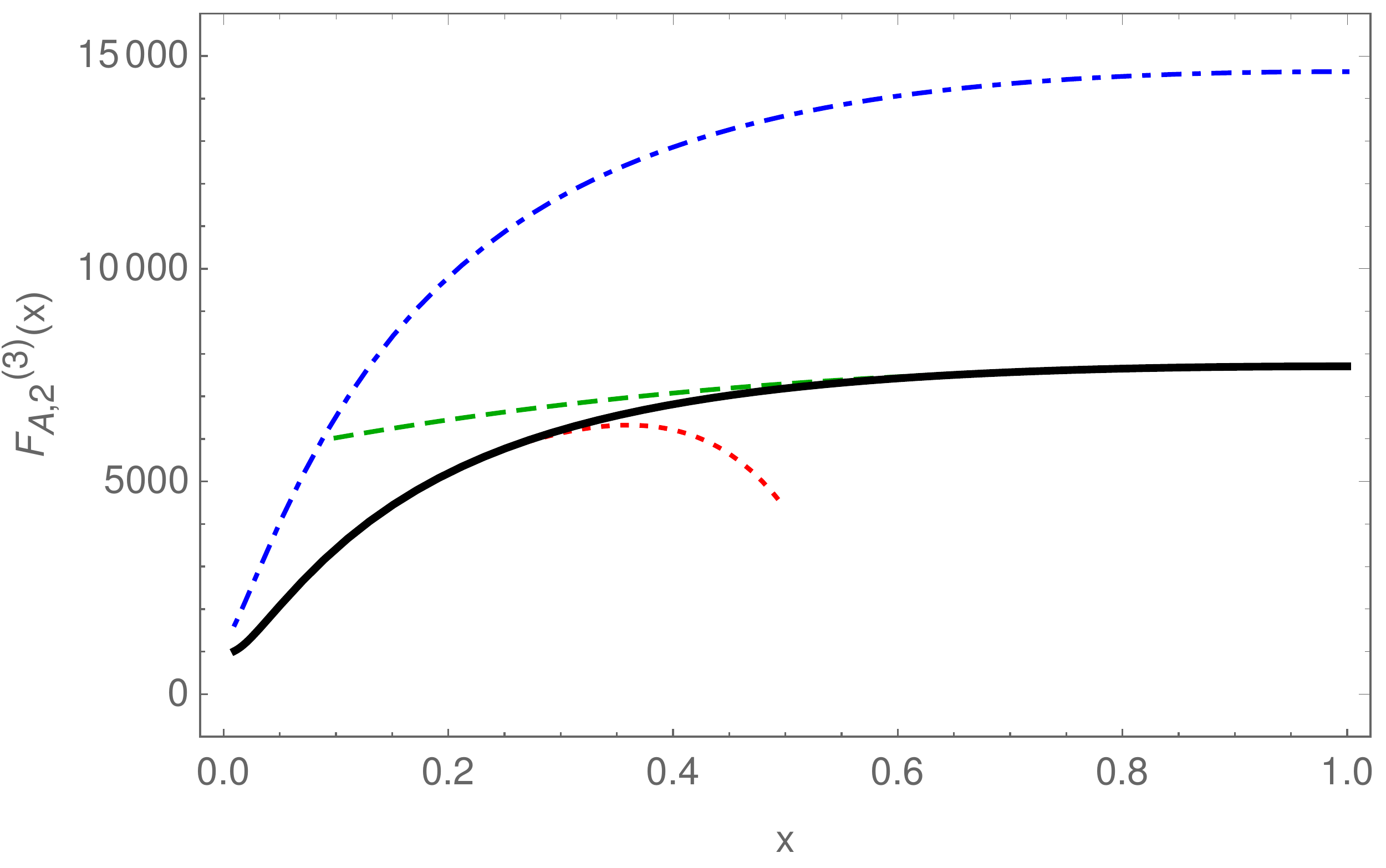}}
\caption{\sf The $O(\varepsilon^0)$ contribution to the vector three-loop form factors 
$F_{V,1}^{(3)}$ (upper left) and $F_{V,2}^{(3)}$ (upper right) and to 
the axial-vector three-loop form factors
$F_{A,1}^{(3)}$ (lower left) and $F_{A,2}^{(3)}$ (lower right)
as a function of $x \in [0,1]$. 
Dash-dotted line: leading color 
contribution of the non--singlet form factor; full line: sum of the complete non--singlet $n_l$-contributions 
for 
$n_l =5$ and the color--planar non--singlet form factor; dashed line: large $x$ expansion; dotted line: small 
$x$
expansion.}
\label{fig:VF12ep0}
\end{figure}

\begin{figure}[ht]
\centerline{%
\includegraphics[width=0.49\textwidth]{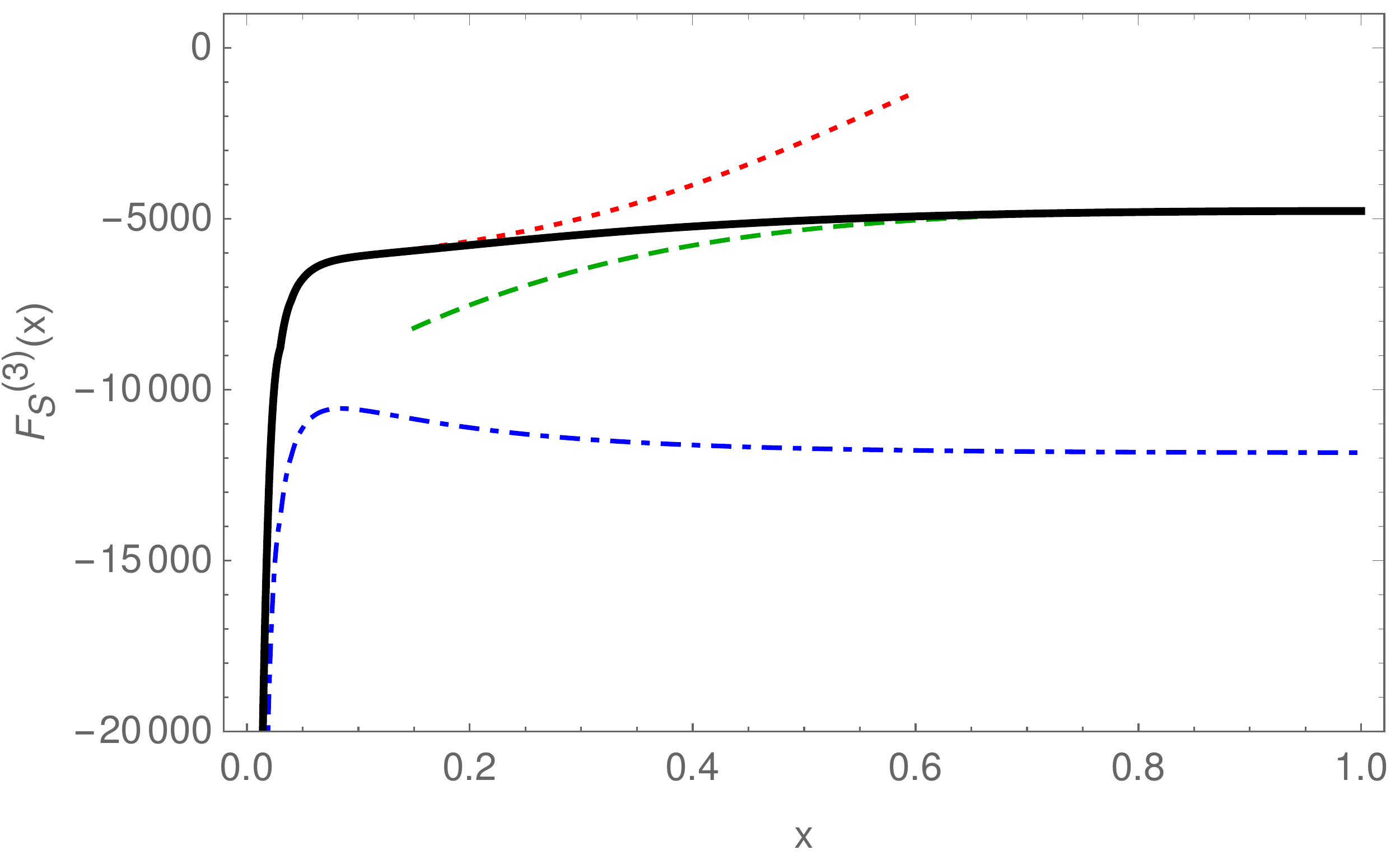}
\includegraphics[width=0.49\textwidth]{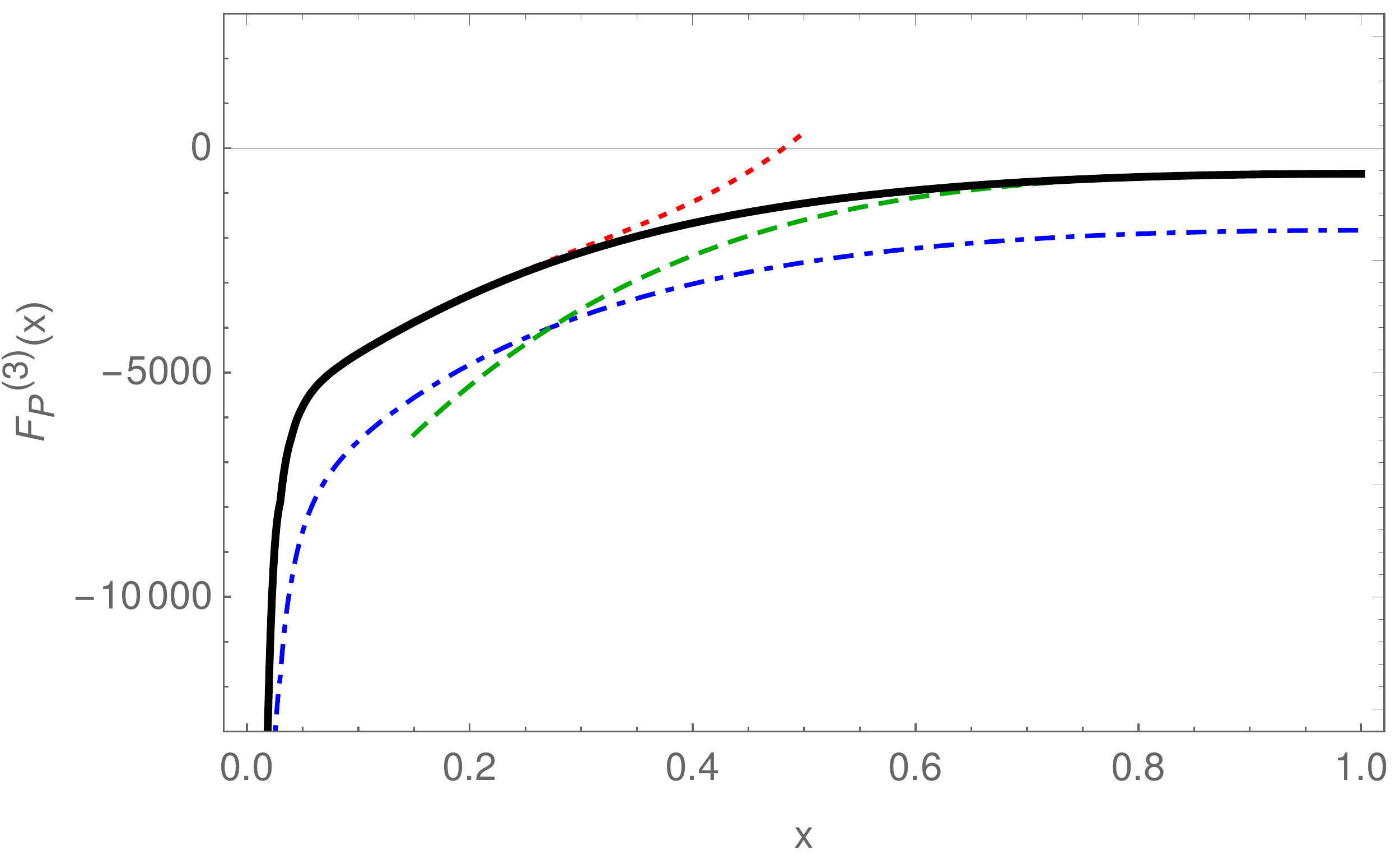}}
\caption{\sf The $O(\varepsilon^0)$ contribution to the scalar three-loop form factors 
$F_{S}^{(3)}$ (left) and to
the pseudo-scalar three-loop form factors
$F_{P}^{(3)}$ (right)
as a function of $x \in [0,1]$. 
Dash-dotted line: leading color 
contribution of the non--singlet form factor; Full line: sum of the complete non--singlet $n_l$-contributions 
for 
$n_l =5$ and the color--planar non--singlet form factor; Dashed line: large $x$ expansion; Dotted line: small 
$x$
expansion.}
\label{fig:SPep0}
\end{figure}

Here we illustrate in Figures~\ref{fig:VF12ep0} the behavior of the $O(\varepsilon^0)$ 
parts of the vector and axial-vector form factors
as a function of $x \in [0,1]$. 
In Figures~\ref{fig:SPep0}, the $O(\varepsilon^0)$ part of the scalar and pseudo-scalar form factors are presented.
We also show their small-- and large--$x$ expansions. The latter 
representations are obtained using {\tt HarmonicSums}. 
To evaluate the HPLs and the cyclotomic HPLs numerically, we use the {\tt GiNaC} package
\cite{Vollinga:2004sn,Bauer:2000cp} and the {\tt FORTRAN}-codes {\tt HPOLY.f} \cite{Ablinger:2018sat} and 
{\tt CPOLY.f} \cite{Ablinger:2018zwz}.

\subsection{Checks}
We have performed a series of checks starting with maintaining the gauge parameter $\xi$ to first order
and thus obtaining a partial check on gauge invariance.
Fulfillment of the chiral Ward identity, Eq.~(\ref{eq:cwi}), gives another strong check on our calculation.
Considering $\alpha_s$--decoupling appropriately, we obtain the universal IR structure for all the UV renormalized 
results, 
confirming again the universality of IR poles. 
Also, in the low energy limit, the magnetic vector form factor produces the anomalous magnetic moment of a heavy quark
which we cross check with \cite{Grozin:2007fh} in this limit.

We have compared our results with those of Ref.~\cite{Henn:2016tyf, Lee:2018nxa, Lee:2018rgs}, which have been 
computed using partly different methods. Both results agree.

\section{Conclusion}
We presented a method to solve uni-variate systems of differential equations which are first order factorizable.
The system is solved in terms of iterative integrals over a finite alphabet of letters.\footnote{The present method has
something in common with the method of hyperlogarithms \cite{Brown:2008um}, which has been used for massive systems in 
\cite{Ablinger:2014yaa} beyond Kummer--Poincar\'e iterated integrals \cite{KUMMER,POINCARE}, but applies to fully 
general alphabets.}
This method can be applied to a wide range of systems in Quantum Field Theory.
We employ this method to solve such a system of differential equations appearing while solving 
the master integrals of the color--planar and complete light quark contributions to the 
three loop heavy quark form factors using the method of differential equations.
Finally, we also obtain all the corresponding form factors for vector, axial--vector, scalar and pseudo--scalar
currents at three loops, which play an important role in phenomenological study of top quark.

\end{document}